# SARF: Enhancing Stock Market Prediction with Sentiment-Augmented Random Forest

Saber Talazadeh[1] and Dragan Peraković [2]

[1] British Columbia Institute of Technology
saber_talazadeh@bcit.ca
[2] University of Zagreb
dragan.perakovic@fpz.hr

**Abstract**

Stock trend forecasting, a challenging problem in the financial domain, involves extensive data and related indicators. Relying solely on empirical analysis often yields unsustainable and ineffective results. Machine learning researchers have demonstrated that the application of random forest algorithm can enhance predictions in this context, playing a crucial auxiliary role in forecasting stock trends. This study introduces a new approach to stock market prediction by integrating sentiment analysis using FinGPT generative AI model with the traditional Random Forest model. The proposed technique aims to optimize the accuracy of stock price forecasts by leveraging the nuanced understanding of financial sentiments provided by FinGPT. We present a new methodology called "Sentiment-Augmented Random Forest" (SARF), which incorporates sentiment features into the Random Forest framework. Our experiments demonstrate that SARF outperforms conventional Random Forest and LSTM models with an average accuracy improvement of 9.23% and lower prediction errors in predicting stock market movements.

**Keywords:** Machine learning, Sentiment Analysis, Finance, Natural Language Processing, Stock Price Prediction, Technical Indicator, Large Language Model, Random Forest

## 1   Introduction

Predicting stock trends is a tough task because of the many factors involved. Despite the development of stock predictors based on statistical models, the dynamic, non-linear, and complex nature of the stock market makes effective trend prediction a persistently challenging task [1]. In the realm of quantitative finance, the focus is on intelligent timing and stock selection. As quantitative investment and machine learning increasingly converge, understanding the rise and fall of stocks becomes pivotal. Diverse stock price forecasting methods exist, each with its own advantages and drawbacks. Machine learning models, in particular, showcase effectiveness and extensibility by learning relationships between predictor variables and stock movement directions in historical data [2]. Unlike traditional statistics and econometric models, machine learning models demonstrate superior prediction performance and robustness. Researchers have explored various machine learning models, such as support vector machines and random forests, for stock trend prediction. Integrating these models presents challenges, especially in handling time series data, selecting technical indicators, and optimizing parameter combinations [3]. This study contributes by systematically building a stock forecasting model that integrates technical indicators with sentiment analysis throughout the process and incorporating exponential smoothing to reprocess technical indicators. The primary contribution of this research is integration of sentiment analysis through the incorporation of sentiment scores and dynamic weight adjustments in the optimized Random Forest model with data sourced from Alpha Vantage. This integration enhances the model's ability to capture information reflecting stock movement and the impact of market sentiment on stock prices.

To extract textual sentiment information, we employ the FinGPT model, a transfer learning model pre-trained on massive finance textual content. This model demonstrates superior performance in finance sentiment analysis. The study's goal is to evaluate the performance of the optimized random forest in medium- and long-term stock forecasting, aiming to improve overall forecasting accuracy. The paper concludes by comparing the prediction performance of SARF, RF, and LSTM based on relevant metrics.

## 2 Related Work

Researchers employ various technologies, including statistics and data mining, to classify and predict future stock values. Tan et al. [8] focus on stock selection, utilizing Chinese stock market data. They combine the fundamental/technical feature space and pure momentum space with a random forest to predict short- and long-term share price trends. Their model achieves a standardized fund performance evaluation index of 2.75 and 5, demonstrating its effectiveness in strategy selection. Kofi et al. [9] explore macroeconomic variables, showing that using more important features to train the random forest model reduces prediction errors by 7.1% compared to models trained with all features. This highlights the positive impact of screening macroeconomic factors on stock market forecasts.

Feature set selection is a critical step in these studies. Ballings et al. [10] investigate traditional and integrated models in machine learning, proving that integrated models outperform single models in predicting financial data based on time series. Random forest with bagging is highlighted as an excellent integrated model, preventing overfitting during training. Basak et al. [11] train random forest and XGBoost using exponential smoothing data, demonstrating increased trend prediction accuracy with an improved time window. Random forest is shown to have more advantages than XGBoost overall. Luckyson et al., relying on technical indicators, use the random forest model to predict stock trends, outperforming support vector machines and logistic regression for more effective trend prediction results.

Yanjun Chen constructs a financial transaction strategy model based on LightGBM to address sparse high-dimensional feature matrices in financial data. The LightGBM model significantly reduces prediction errors and achieves higher prediction precision compared to OpenGL Mathematics, deep neural networks, and support vector machines [15]. SVM, as studied by Manik et al., incorporates structured risk minimization to decrease errors and improve classification effectiveness. In their study, intraday stock status is mined using various classifiers, including C4.5, random forest, logistic regression, linear discriminant, SVM, quadratic SVM, cubic SVM, Gaussian SVM, and others. The performance of different classifiers is evaluated based on accuracy, misclassification rate, precision, recall, and other metrics [12]. Decision trees, particularly effective for discrete features, demonstrate superior performance in certain scenarios [13].

## 3    Proposed Method:

In this section, we propose a method that leverages the Random Forest model to integrate technical indicators with sentiment analysis using FinGPT. We extract sentiment scores (positive, negative, neutral) for each data point. By incorporating sentiment-based features extracted from FinGPT using financial news articles and using the sentiment scores as additional features, we enhance the Random Forest model's capacity to capture and incorporate market sentiment [19].

### 3.1    Random Forest Model:

We utilized the Random Forest algorithm as the foundational model, leveraging its robustness and capacity to manage relationships within financial data through a robust ensemble learning approach. This algorithm proved effective in handling non-linear relationships and mitigating overfitting challenges present in financial datasets. The model's capability to furnish feature importance scores played important role in discerning the significant factors influencing stock movements [18]. Among the critical hyperparameters applied in our research are number of trees, maximum tree depth, minimum samples required for splitting and the feature subset size. The tuning of these hyperparameters was used for optimizing the predictive performance of the model within the dynamic nature of stock markets.

### 3.2    Sentiment Analysis with FinGPT:

In our study, we conducted experiments with both the FinGPT and FinBert models for sentiment analysis. FinGPT demonstrated notable strengths, particularly in its capacity to comprehend context, generate coherent responses, and provide diverse financial insights beyond sentiment analysis. Due to its broader focus, adeptness in handling various financial queries, and proficiency in generating responses in natural language, we opted to utilize FinGPT in our research. FinGPT is utilized to perform sentiment analysis on financial news. The model's contextual understanding of financial language provides valuable sentiment scores. It is specifically designed to provide information and answer questions related to finance, banking, and investing. FinGPT uses natural language processing (NLP) technology to understand and respond to user queries, and it has been trained on a large dataset of financial information to ensure that its responses are accurate and relevant.

It also provides a built-in sentiment analysis API that can be used to analyze text data and extract sentiment scores. In this study we used FinGPT APIs to analyze text data and get a sentiment score ranging from -1 (negative) to 1 (positive).

### 3.3 SARF - Sentiment-Augmented Random Forest:

We introduced the SARF model, which combines technical indicators and sentiment features with the Random Forest model. We added sentiment-based features from FinGPT as extra inputs to improve the Random Forest model's ability to grasp market sentiment. This hybrid approach aims to leverage the complementary strengths of both models. SARF builds upon the ensemble learning paradigm of Random Forests by introducing a mechanism to integrate sentiment analysis and technical indicators. The SARF model consists of an ensemble of decision trees, each trained on different subsets of the dataset, but with the inclusion of sentiment-based and technical indicator features.

SARF builds upon the ensemble learning paradigm of Random Forests by introducing a mechanism to integrate sentiment analysis and technical indicators. The SARF model consists of an ensemble of decision trees, each trained on different subsets of the dataset, but with the inclusion of sentiment-based and technical indicator features.

We used TA-Lib (Technical Analysis Library) to calculate 15 technical indicators. During feature selection, variables with high correlation are eliminated to addresses multicollinearity issues, and helps prevent overfitting. This process enhances the model's interpretability, improves computational efficiency, and aids in dimensionality reduction, making the model more practical and resource-efficient. Subsequently, the selected indicators serve as input vectors for training the SARF model, and the model's performance is assessed on the test dataset.

**Figure 1**. Data Processing Diagram

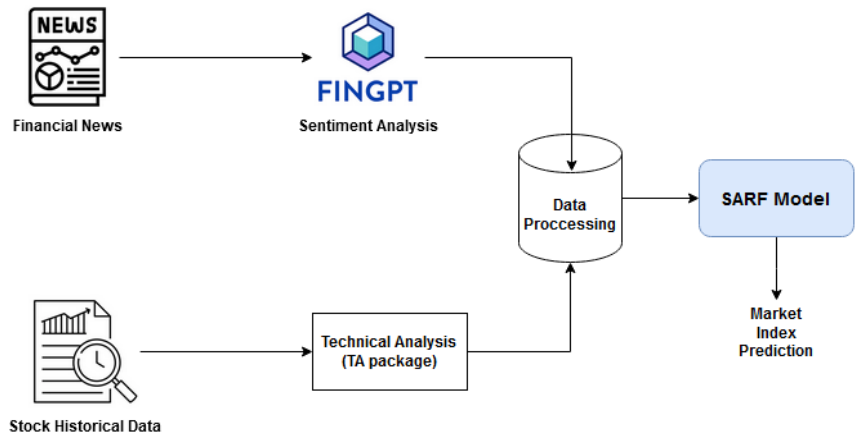

## 4 Experimental Evaluation

We conduct extensive experiments using historical financial data, comparing the predictive performance of SARF against traditional Random Forest models. Evaluation metrics include accuracy, precision, recall, and F1 score. We utilized cross-validation to ensure robustness and minimize overfitting in the proposed model.

### 4.1 Data collection

For data collection we used Alpha Vantage for stock market and technical indicator by calling their APIs. Alpha Vantage is known for offering a wide range of financial data APIs, including stock market data, technical indicators, and historical prices. This study utilizes data on the price and volume1 of market indices for testing. NASDAQ, S&P 500, and Dow Jones are prominent stock market indices that respectively represent technology stocks, a broad market cross-section, and 30 major industrial companies, reflecting the overall performance of the U.S. stock market.

We chose to use US market indices such as NASDAQ, S&P 500, and Dow Jones for stock market predictions instead of individual stocks because these indices give a broader and more varied view compared to looking at single stock- of a company. They include a mix of different parts of the market, covering various industries and companies, making them a stronger reflection of overall market trends. Predicting the market based on these indices let us consider a wider range of factors, like big economic changes, and reduces the impact of events specific to one company that might affect its stock. Plus, these indices have a stable history, making them helpful for understanding overall stock market performance and trend.

The dataset was collected at daily intervals by querying Alpha Vantage APIs, capturing key metrics such as opening price, lowest price, highest price, closing price, and trading volume. The data spanned from January 2, 2015, to December 30, 2023.
In this study, we leveraged these technical indicators as independent variables to predict future stock market movements. Technical indicators are mathematical calculations derived from historical data, providing insights into trading patterns for financial assets. Throughout the study, we utilized several commonly used indicators, some of which have been previously explored by other researchers.

The learning algorithm used in our paper is random forest. The time series data is acquired, smoothed and technical indicators are extracted as shown in table 1. Technical indicators are parameters which provide insights to the expected stock price behavior in future. These technical indicators are then used to train the random forest. The time series historical stock data is first exponentially smoothed. Exponential smoothing applies more weightage to the recent observation and exponentially decreasing weights to past observations.

**Table 1.** Technical Indicators

| Indicator Name | Description |
| --- | --- |
| **Moving Averages (MA)** | The average value of a security over a given time. Helps identify trends and potential reversals. |
| **Moving Average Convergence Divergence (MACD)** | Measures the relationship between two moving averages. Signals trend strength and direction. |
| **Relative Strength Index (RSI)** | Measures the speed and change of price movements. Indicates overbought or oversold conditions. |
| **Stochastic Oscillator** | Compares a security's closing price to its price range over a specific period. Shows momentum. |
| **Williams %R** | Measures overbought or oversold levels. Similar to the stochastic oscillator. |
| **Bollinger Bands** | Consists of three lines: moving average, upper band, and lower band. Indicates volatility and trends. |
| **On-Balance Volume (OBV)** | Measures positive and negative volume flow. Helps predict price movements. |
| **Accumulation / Distribution Line (ADL)** | Tracks buying and selling pressure. Reflects accumulation or distribution of a security. |
| **Aroon Oscillator** | Identifies the strength and direction of a trend. Combines Aroon Up and Aroon Down indicators. |
| **Average True Range (ATR)** | Measures market volatility. Indicates potential price movement. |
| **Ichimoku Cloud** | Combines several indicators to provide a comprehensive view of support, resistance, and trends. |
| **Parabolic SAR (Stop and Reverse)** | Helps identify potential reversal points. Useful for setting stop-loss orders. |
| **Fibonacci Retracement** | Uses Fibonacci ratios to predict potential retracement levels in price movements. |
| **Chaikin Money Flow (CMF)** | Combines price and volume data to assess buying and selling pressure. |
| **Average Directional Index (ADX)** | Measures trend strength. Helps determine whether a security is trending or ranging. |

### 4.2 Feature extraction

In the area of technical analysis, an important parameter is the technical index derived from stock data, serving as a predictor for stock market trends and a tool frequently utilized by investors. These indicators play a significant role in evaluating short-term stock price dynamics and can prove effective for medium- and long-term purposes, such as identifying entry and exit points. To validate the efficacy of the process-optimized stochastic forest model proposed in this study, we employ key technical indicators and sentiment indicators from FinGPT as input features for model training. The output variable is whether the market index moves up or down, predicting the dynamic trend of the stock market index. Initially, various indicators for technical analysis are selected based on the objective of medium- and long-term stock forecasting.

To establish an effective technical data system, we carefully select three indices as the dataset and conduct feature importance analysis by constructing three decision trees. The resulting technical feature importance data aids in the selection of appropriate indicators. The overall dataset of technical indicators is then divided into a training set (two-thirds) and a test set (one-third). From a combination of 15 technical indicators and 4 sentiment indicators, we derive 14 predictor variables for forecasting. Through exploratory analysis and correlation coefficient calculations, it is evident that some predictor variables show correlations. The precise correlations or highly correlated relationships can lead to multicollinearity, impacting model stability or hindering accurate parameter estimation. To address this, the SARF algorithm is employed, introducing a penalized function to compress less important variables and eliminate multicollinearity.

Correlation analysis is performed to calculate the correlation coefficients between predictor variables, and highly correlated features (typically exceeding 0.8 correlation) are eliminated to prevent redundancy. This strategic pruning ensures our model remains efficient and does not process redundant information [24].

To further mitigate multicollinearity, principal component analysis (PCA) and ridge regression are employed. Ridge regression, a penalized regression method, effectively shrinks the coefficients of highly correlated features towards zero, mitigating the multicollinearity issue and enhancing the stability of the model [25].

### 4.3 Parameter optimization

In our efforts to optimize the Random Forest model using parameter optimization, we applied various techniques. We began with Super parametric optimization, adjusting the number of trees (n_estimator) for the S&P index manually and evaluating precision, recall rate, F1 value, and accuracy rate. We then explored Grid Search, systematically searching for the best parameter values within a specified range. However, due to its inefficiency and resource consumption, we turned to Random Search, a more valuable and efficient alternative based on low effective dimension. We optimized parameters like the number of decision trees, maximum tree depth, and minimum samples for node subdivision and leaf nodes. Model performance was evaluated through 3-fold cross-validation, using the AUC score of the ROC curve as the primary index. We also considered a fixed random number seed to mitigate sampling error. We observed that more trees in the Random Forest model don't necessarily mean better performance, considering resources, stability, and the model's robustness to outliers. Optimal parameters, obtained through random sampling, were evaluated using the AUC score and demonstrated effectiveness by comparing results before and after optimization.

### 4.4     Evaluating indicator

Our assessment goes beyond simple metrics like accuracy, delving deeper into various performance indicators specific to the chosen classification model. Accuracy, while helpful, doesn't tell the whole story. We delve into precision, representing the accuracy of positive predictions, and recall, revealing the model's ability to capture true positives. We further utilize the F1-score, a harmonic mean of precision and recall, providing a balanced measure of overall effectiveness. These metrics offer a nuanced understanding of the model's strengths and weaknesses. In our analysis, we focused on imbalanced classes were present, and we found that metrics such as AUC-ROC and precision-recall curves offered valuable insights

Through a comprehensive evaluation, we gained a clear picture of the model's performance, enabling us to make informed decisions about its suitability for real-world deployment. We identified potential biases, areas for improvement, and opportunities for further optimization, ultimately ensuring the model effectively addressed the target problem.

## 5     Experimental Result

Preliminary results demonstrate that SARF outperforms conventional Random Forest models, showcasing its efficacy in predicting stock market movements, particularly in volatile market conditions. Our study revealed evidence of the better performance of the SARF model compared to conventional Random Forest models. Through parameter optimization, we identified an optimal combination of parameters, leveraging different time windows for predicting stock market movements. Specifically, the model's precision, recall rate, F1 value, and accuracy rate were evaluated at a 60-day time window, demonstrating its efficacy for medium and long-term predictions within the 62–82 days range. These preliminary findings underscore the potential of SARF, particularly in volatile market conditions.

In the comparative analysis with the LSTM model, our constructed optimized random forest stock prediction model exhibited advantages. ROC curves drawn from the training results showcased the performance of each model, with the comprehensive results presented in Table 2 indicating the optimized random forest model's superior performance over the original random forest and LSTM models. This emphasizes the generalization ability and superior trend prediction capability achieved through the random forest algorithm's decision tree feature extraction and random parameter optimization. The utilization of Precision-Recall curves further deepened our understanding, highlighting the SARF model's stability in stock forecasting, especially when compared to the LSTM model across various stock market index. These findings substantiate the effectiveness of SARF in providing robust predictions in dynamic stock market environments. The hybrid nature of SARF offers some advantages over traditional Random Forest models such as inclusion of sentiment features and technical indicators provides additional insights into market sentiment and historical price movements, allowing for a more comprehensive understanding of the model's predictions.

**Table 2.** Performance Comparison

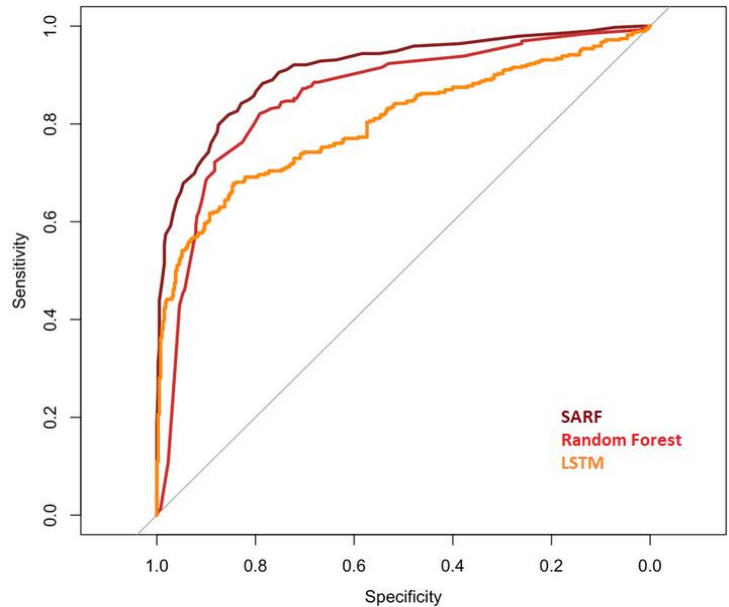

## 6  Conclusion

This research introduces a new technique, SARF, for optimizing the accuracy of stock market predictions by combining sentiment analysis with FinGPT and optimized

Random Forest model. The promising results indicate the potential of this approach for real-world applications in financial forecasting.

Table 3. Experiments Outcome evaluating the accuracy of models on stock indices

| Index | Traditional Random Forest | LSTM | Optimized Random Forest (SARF) |
|---|---|---|---|
| S&P 500 | 0.67 | 0.58 | 0.78 |
| Nasdaq | 0.64 | 0.69 | 0.85 |
| Dow Jones | 0.59 | 0.61 | 0.82 |

# 7     Future Work

Future research will explore the scalability of SARF to handle larger datasets, investigate additional sentiment features using other LLM in financial domain, and assess its performance in diverse market conditions. Additionally, the integration of real-time sentiment analysis could further enhance the model's responsiveness to dynamic market changes. It would be worth using ML optimization techniques to tune the hyperparameters and improve the model prediction accuracy further.